\documentclass[aps,prl,reprint,groupedaddress]{revtex4-1}
\usepackage{amsmath} 
\usepackage{amsthm} 
\usepackage{amssymb}	
\usepackage{graphicx} 

\usepackage[T1]{fontenc}
\usepackage{units}
\usepackage{bbold}
\begin{document}


\title{Optical evidence of broken C$_4$ symmetry across optimal doping in BaFe$_{2}$(As$_{1-x}$P$_x$)$_2$}


\author{E. Thewalt}
\affiliation{Materials Science Division, Lawrence Berkeley National Laboratory, Berkeley, California 94720, USA}
\affiliation{Department of Physics, University of California, Berkeley, California 94720, USA}

\author{J. P. Hinton}
\affiliation{Materials Science Division, Lawrence Berkeley National Laboratory, Berkeley, California 94720, USA}
\affiliation{Department of Physics, University of California, Berkeley, California 94720, USA}

\author{I. Hayes}
\affiliation{Materials Science Division, Lawrence Berkeley National Laboratory, Berkeley, California 94720, USA}
\affiliation{Department of Physics, University of California, Berkeley, California 94720, USA}

\author{T. Helm}
\affiliation{Materials Science Division, Lawrence Berkeley National Laboratory, Berkeley, California 94720, USA}
\affiliation{Department of Physics, University of California, Berkeley, California 94720, USA}

\author{D. H. Lee}
\affiliation{Materials Science Division, Lawrence Berkeley National Laboratory, Berkeley, California 94720, USA}
\affiliation{Department of Physics, University of California, Berkeley, California 94720, USA}

\author{James G. Analytis}
\affiliation{Materials Science Division, Lawrence Berkeley National Laboratory, Berkeley, California 94720, USA}
\affiliation{Department of Physics, University of California, Berkeley, California 94720, USA}

\author{J. Orenstein}
\affiliation{Materials Science Division, Lawrence Berkeley National Laboratory, Berkeley, California 94720, USA}
\affiliation{Department of Physics, University of California, Berkeley, California 94720, USA}


\date{\today}

\begin{abstract}
The isovalently doped high-$T_{\text{c}}$ superconductor BaFe$_2$(As$_{1-x}$P$_x$)$_2$ is characterized by a rich temperature-doping phase diagram, which includes structural, antiferromagnetic, and superconducting phase transitions. In this work, we use photomodulated reflectance at 1.5 eV to detect the onset of linear birefringence in the normal state, and by inference, the breaking of four-fold rotational (C$_4$) symmetry. We find that C$_4$ breaking takes place in the normal state over a range of $x$ that spans the optimal P concentration, unlike the line of orthorhombic-tetragonal and magnetic phase transitions, which terminates at, or just below, the optimal value of $x$.
\end{abstract}


\maketitle





The discovery of superconductivity in the iron pnictides~\cite{kamihara2008iron,takahashi2008superconductivity,rotter2008superconductivity,paglione2010high} has opened promising new directions in the effort to fully understand the phenomenon of high-$T_\text{c}$ superconductivity.  Of particular note is the possible close connection between superconductivity and nematicity in both the iron pnictide and the copper oxide materials~\cite{fradkin2010nematic}.  In the context of electronic properties of crystalline systems, nematicity denotes the lowering of four-fold to two-fold rotational symmetry C$_4\rightarrow$~C$_2$ without loss of discrete translational symmetry~\cite{kivelson2008iron}.  Hope abounds that the development of a clear picture of the interplay between superconductivity and nematicity will yield further insight into the high-$T_\text{c}$ problem.

The BaFe$_2$(As$_{1-x}$P$_x$)$_2$ (P:Ba122) system has been the focus of significant attention because isovalent doping, that is, substitution of P for As, generates an iron-pnictide superconductor with less disorder than doping on the Fe site. Moreover, evidence for a nematic quantum critical point in P:Ba122 near optimal doping has been reported~\cite{shibauchi2014quantum}.  A quantum critical point is defined by a continuous phase transition at zero temperature, requiring in this system a characteristic line of true symmetry breaking to terminate beneath the superconducting dome.  In P:Ba122, C$_4$ symmetry is apparently broken along a line of simultaneous spin-density-wave and tetragonal-to-orthorhombic structural transitions, $T_s(x)$, that penetrates the superconducting dome just below optimal doping~\cite{nakai2010unconventional,chu2010plane,chu2012divergent,kuo2012magnetoelastically,allred2014coincident}.  While the intercept of $T_s(x)$ with the $T=0$ axis is a candidate nematic quantum critical point, torque magnetometry measurements complicate this picture by suggesting that C$_4$ symmetry is broken along a line, $T_\text{nem}(x)$ that extends beyond optimal doping and well into the overdoped regime~\cite{kasahara2012electronic}.

Clearly, C$_4$ symmetry cannot be broken first at $T_\text{nem}$ and again at $T_s$.  One resolution of this apparent paradox is to suggest that $T_\text{nem}$ marks the onset of strong nematic fluctuations, which couple with residual uniaxial strain in the crystal to mimic spontaneous symmetry breaking.  In this case the order parameter would be expected to show a gradual onset with decreasing $T$, following the Curie-like nematic susceptibility~\cite{chu2012divergent}.  However, a second possibility is that C$_4$ is indeed broken at $T_\text{nem}$, but that the crystal lattice interacts very weakly with nematic order, such that the amplitude of the orthorhombic lattice distortion is below the detection limit.  A realization of this possibility in a different materials system is the breaking of C$_4$ symmetry in Sr$_3$Ru$_2$O$_7$~\cite{borzi2007formation}, where the accompanying lattice distortion is more than 100 times smaller than occurs below $T_s$ in iron pnictides~\cite{stingl2011symmetry}. A distinguishing feature of this scenario is that the onset of symmetry breaking would be abrupt rather than following a power law in $T-T_s$. Here we report a sharp onset of optical birefringence at temperatures above $T_s(x)$, indicating static long-range C$_4$ symmetry breaking along a line in the phase diagram that extends across the dome of superconductivity and is distinct from the structural/magnetic phase boundary that terminates near optimal doping.

The optical birefringence is observed by a form of modulation spectroscopy in which a weak laser pulse perturbs the equilibrium reflectance $R$ and the subsequent change in reflectance, $\Delta R(t)$, is measured by a time-delayed probe pulse.  Although this pump/probe technique is often applied to measurements of the lifetime of photoinduced quasiparticles~\cite{gedik2004single,talbayev2012relaxation,torchinsky,mihailovic}, in this work we focus on the amplitude of $\Delta R$ near zero time-delay, emphasizing its dependence on temperature and polarization of the probe.  Detecting the photomodulated reflectance is closely related to thermo- and electroreflectance spectroscopy, techniques that have long been used to enhance weak structure in $R$~\cite{cardona1967electroreflectance}.  In the case of photomodulated reflectance, the perturbation is a scalar quantity (the energy, $u$, absorbed from the pump pulse), so $\Delta R$ is a rank two response tensor like $R$ itself.

Our initial measurements of $\Delta R$ on the P:Ba122 system revealed a strong dependence of its amplitude, and even sign, on the position of the laser beam with respect to the sample. To begin to characterize the spatial inhomogeneity we registered each position on a sample with respect to an optical landmark on the sample mount using a high-resolution video feed, allowing us to define beam position to a precision of $5$~$\mu$m, which is small compared to the $50~\mu$m beam spot.  We then used this optical landmarking capability to generate maps of the spatial variation of $\Delta R$. The measurements were performed using 100~fs pulses from a mode-locked Ti:Sapphire laser at 80~MHz repetition rate, 800~nm center wavelength, and $1~\mu$J$/$cm$^2$ fluence.

\begin{figure}[h!]
\includegraphics[]{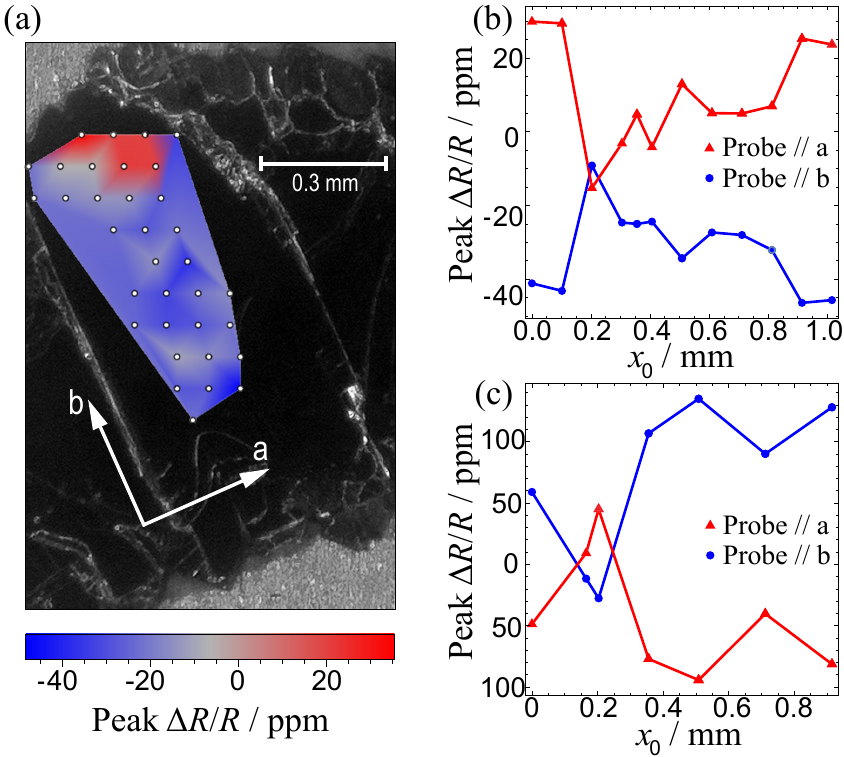}
\caption{Spatial variation of $\Delta R/R$ in the underdoped ($x = 0.24$) and optimal ($x = 0.31$) samples.  In (a), the underdoped sample is pictured with orthorhombic principal axes and scale indicated in white, and a color map of  $\Delta R/R$ at 6~K is overlaid.  Open circles indicate points where data were collected.  The sign of $\Delta R$ is opposite in the red and blue regions, suggesting that they are preferentially populated with oppositely-oriented C$_4$-breaking domains.  Panels (b) and (c) show $\Delta R/R$ as a function of position $x_0$ along a line in the optimal sample at $22$~K~(b) and $6$~K~(c) and with probe along the a-axis (red triangles) and b-axis (blue circles).  Again, the variation in signal strength and sign indicates changes in the relative population of a- and b-oriented domains in the beam spot.\label{fg:domains}}
\end{figure}

Figure~\ref{fg:domains}(a) shows a picture of an underdoped ($x = 0.24$) sample overlaid with a false-color map of the amplitude of $\Delta R/R$ near zero time delay, measured with polarization of the probe beam parallel to one of the Fe-Fe bond directions that define the orthorhombic principal axes.  We refer to these axes, for brevity, as the \textbf{a} and \textbf{b} axes.  The contrast revealed by linear polarization of the probe suggests that the origin of the spatially inhomogeneous response is the presence of birefringent domains.

Figs.~\ref{fg:domains}(b) and (c) show line cuts of $\Delta R/R$ maps for probe polarization parallel to the \textbf{a} and \textbf{b} axes, measured at $T=$ 22 K and 6 K, respectively, for an optimally doped ($x = 0.31$) sample.  Here the photomodulated reflectance indicates strong birefringence in a sample that does not exhibit a structural transition in the normal state. The scans reveal another surprising feature: \textit{the birefringence of} $\Delta R/R$ \textit{manifests as sign reversal under a rotation of the linear polarization of the probe light from the} $\textbf{a}$ \textit{to the} $\textbf{b}$ \textit{crystallographic direction}.  By measuring $\Delta R/R$ as a function of the direction of linear probe polarization, we have confirmed that the principal optical axes correspond with the Fe-Fe bond direction within 2$^\circ$.

The observation that $\Delta R$ changes sign under $\pi/2$ rotation of the probe polarization greatly clarifies the origin of the observed spatial inhomogeneity.  We infer that regions of the sample where $\Delta R$ is relatively small are those in which many orthogonally oriented domains lie within the focal spot of the probe.  Conversely, areas with large and relatively homogeneous $\Delta R$ correspond either to regions with predominately one of the two possible orientations.  We note that, overall, similar phenomena were observed in a detailed study of $\Delta R$ in the Ba(Fe$_{1-x}$Co$_x$)$_2$As$_2$~\cite{mihailovic2}, although, as we detail below, the $T$ dependence of birefringence in the two systems appears to be different.

\begin{figure}
\includegraphics[]{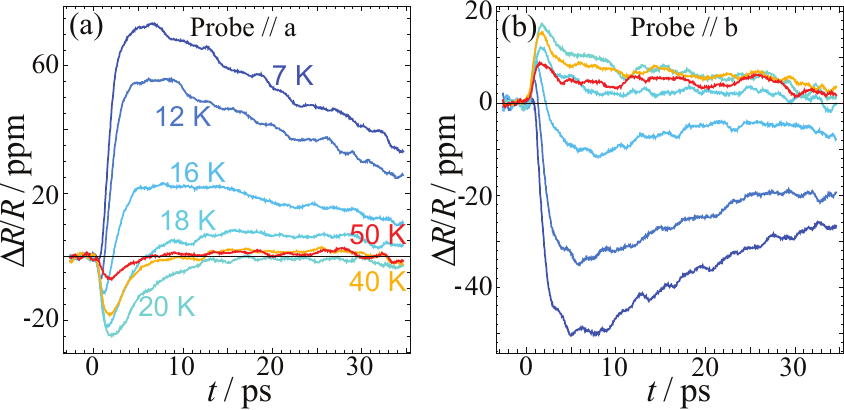}
\caption{Transient $\Delta R/R$ response in the optimally doped ($x = 0.31$) sample with probe along a-axis (a) and b-axis (b), shown at several representative temperatures labeled in (a).  The response along the a-axis is opposite in sign to that along the b-axis.  At intermediate temperatures, the response has a strong two-component character.\label{fg:repcurves}}
\end{figure}

To map the time and temperature dependence of the single domain response we used our landmarking capability to fix the probe beam on a spot in a homogenous region as the sample is cooled. Figure~\ref{fg:repcurves} shows traces of $\Delta R/R$ \textit{vs}.\ time delay, $t$, measured on an optimally doped ($x = 0.31$) sample for several temperatures, while maintaining the sample position fixed as described above. These data reveal that $\Delta R(t)/R$ comprises two distinct components that decay on different timescales, both of which change sign with rotation of probe polarization.

\begin{figure}
\includegraphics[]{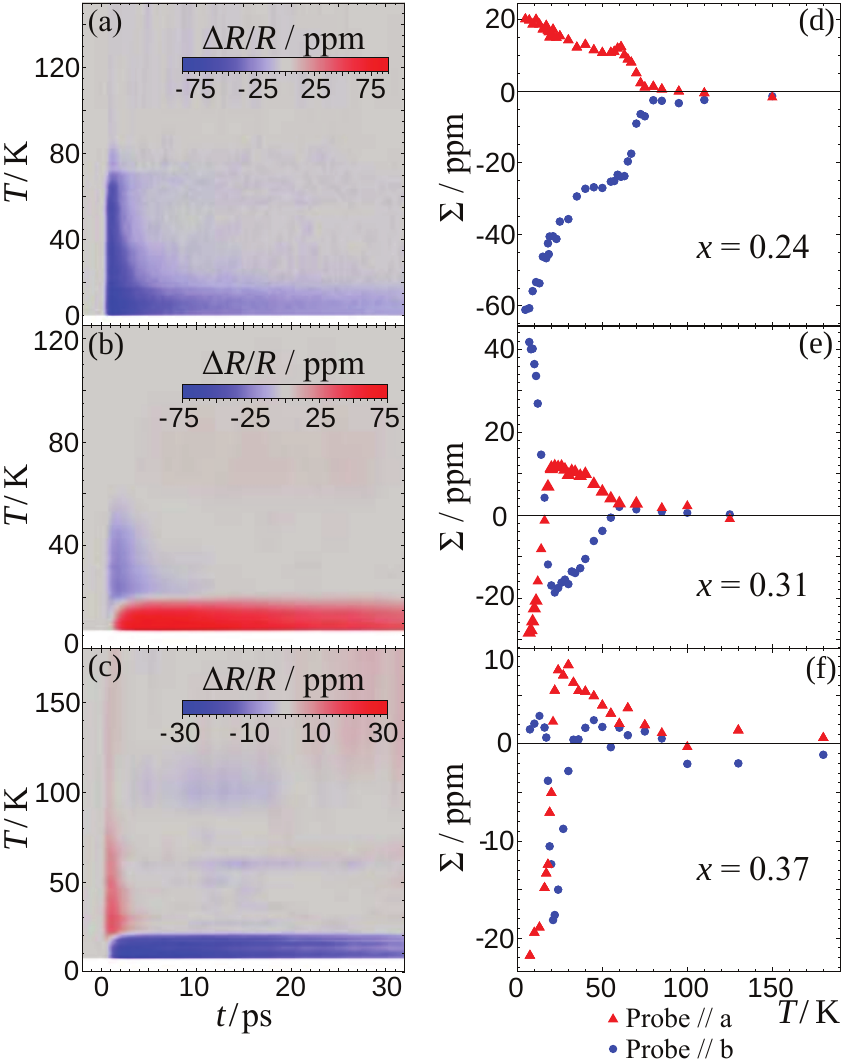}
\caption{Transient $\Delta R/R$ response in underdoped ($x=0.24$), optimally doped ($x=0.31$), and overdoped ($x=0.37$) BaFe$_2$(As$_{1-x}$P$_x$)$_2$.  (a-c) False color plots of the time and fixed-spot temperature dependence of $\Delta R/R$.  (d-e) Fixed-spot temperature dependence of $\Delta R/R$ averaged over the time delay window $0<t<5$ ps with probe along a-axis (red triangles) and b-axis (blue circles).\label{fg:drrdata}}
\end{figure}

An overview of the dependence of the two components of $\Delta R/R$ on time and temperature is presented using false-color images in the first column of Fig.~\ref{fg:drrdata}.  The three rows correspond to underdoped, optimal, and overdoped samples ($x=0.24$, 0.31, and 0.37, respectively).  The temperature dependence was measured while maintaining the laser spot at a fixed position in a homogeneous region of the sample.  The color plots help to visualize the main features of the modulation in reflectance: (i) $\Delta R$ is below our detection sensitivity at room temperature, (ii) a single, short-lived component appears upon cooling at a well-defined temperature in the normal state, and (iii) an additional, long-lived component appears upon entering the superconducting state.

The second column of Fig.~\ref{fg:drrdata} contrasts the temperature dependence of $\Delta R$ for two orthogonal linear polarizations of the probe. To enhance signal-to-noise ratio, we plot $\Sigma(T)$, defined as $\Delta R(t,T)/R$ averaged over $t$ in the interval from 0 to 5~ps.  For all samples, the photomodulation of $\Delta R$ appears abruptly in the normal state.  In the $x = 0.24$ and $x = 0.31$ samples, as soon as $\Delta R$ appears it changes sign under $\pi/2$ rotation of the polarization of the probe; that is, there is no range of $T$ where a C$_4$ symmetric $\Delta R$ is observed.  The second component that emerges near the superconducting transition temperature has the same sign as the normal state component in the underdoped sample, but the opposite sign in the optimal and overdoped samples, a feature of the data that we return to below.

Next we address two questions raised by the measurements described above: (1) why does the breaking of C$_4$ symmetry manifest as a sign reversal of $\Delta R(T)/R$ under $\pi/2$ rotation of probe polarization, and (2) why do the onset temperature of nonzero $\Delta R/R$ and its sign reversal property coincide in the underdoped and optimally doped samples?  Photomodulation of $R$ can occur in any material when optical excitation perturbs the single particle Fermi-Dirac or Bose occupation factors, causing a change in the optical conductivity or dielectric function.  However, in systems that are near a phase transition, a distinct source of $\Delta R$ is photoinduced weakening of an order parameter, as has been demonstrated in a wide variety of superconductors, magnets, and charge-density wave systems~\cite{orenstein2012ultrafast}.  From the sign reversal property of $\Delta R/R$ described above, we infer that weakening of an order parameter that breaks C$_4$ symmetry is the source of the photomodulation of $R$.

Below we describe more quantitatively the main features of $\Delta R$ that follow from photomodulation of an order parameter that breaks C$_4$ symmetry.  We assume for the sake of specificity that the symmetry breaking is a manifestation of orbital order, that is, a splitting of the bands originating from Fe $d_{xz}$ and $d_{yz}$ orbitals, as has been seen by ARPES~\cite{zxshen}.  At optical frequency $\omega$ the equilibrium optical reflection amplitude $r(\omega)$ will have a contribution, $r_d(\omega)$, from $d_{xz}$ and $d_{yz}$ states near the Fermi level, in addition to a background contribution from other bands, $r_b(\omega)$, such that $r(\omega)=r_b(\omega)+r_d(\omega)$. If orbital order induces a splitting of the $d$-related bands by $\hbar\Omega$, their contribution will undergo a polarization-dependent shift, where, for example $r_{dx}(\omega)=r_d(\omega+\Omega/2)$ and $r_{dy}(\omega)=r_d(\omega-\Omega/2)$.

While the change in $r_d(\omega)$ that results from this shift can be difficult to observe, particularly when the spectra are broad, modulation spectroscopy can enhance the order parameter-related features relative to the background.  We assume (1) that the fractional change in the order parameter, $-\delta \Phi/\Phi$, caused by photoexcitation is proportional to the absorbed pump photon energy $u$, and (2) that $\Omega\propto\Phi$. We parameterize these proportional relations by writing $-\delta \Phi/\Phi=\alpha u$ and $\Omega(T)=\beta\Phi(T)$. Under these assumptions, the photomodulation of the reflection amplitude is given by
\begin{equation}
\frac{\partial r_{dx,dy}(\omega)}{\partial u}=(+,-)\alpha\beta\left(\frac{\partial r_d(\omega)}{\partial\Omega}\right)\Phi(T).
\label{eq:drdu}
\end{equation}
Equation~\ref{eq:drdu} shows that photomodulation of the splitting of bands associated with the $d_{xz}$ and $d_{yz}$ orbitals accounts for the sign-changing property of $\Delta R$. The absolute magnitude of $\Delta R$ will be equal for the two polarizations if the shift of the bands is equal and opposite, as was assumed for simplicity above. In general $|\Delta R|$ will be different for the two polarizations, as is the case for the data shown in Fig.~\ref{fg:drrdata}. Note that Eq.~\ref{eq:drdu} predicts that the modulated reflectance is directly proportional to the C$_4$ breaking order parameter.

Another striking feature of the data is that, for the samples with $x=0.24$ and $x=0.31$, $|\Delta R(T)|$ for the two orthogonal polarizations are proportional over the entire $T$ range, despite the fact that there are two components arising from distinct phases. This observation strongly suggests that $\Delta R$ arises from photomodulation of the same band splitting $\Omega$ in both the normal and superconducting states. If the photomodulation of the superconducting order, $\Psi$, obeys $\delta\Psi/\Psi=-\alpha_\Psi u$, then a coupling of the form $\Omega(T)=\beta_\Psi\Psi(T)$ yields an overall temperature dependence
\begin{equation}
\Delta R_{a,b}(T) \propto (+,-)\left(\frac{\partial r_d(\omega)}{\partial\Omega}\right) [\alpha\beta\Phi(T)+\alpha_\Psi\beta_\Psi\Psi(T)],
\label{eq:drdu2}
\end{equation}
which has the property that $\Delta R$ is proportional for the two polarizations. Note that a nonzero $\beta_\Psi$ coefficient arises naturally when the C$_4$ breaking and superconducting orders are coupled, such that $\delta \Psi \propto \delta \Phi$. The observation, mentioned above, that the relative sign of the two components of $\Delta R$ changes with $x$ can be explained by assuming that the interaction of the two forms of order is attractive below optimal doping and repulsive above.

\begin{figure}
\includegraphics[]{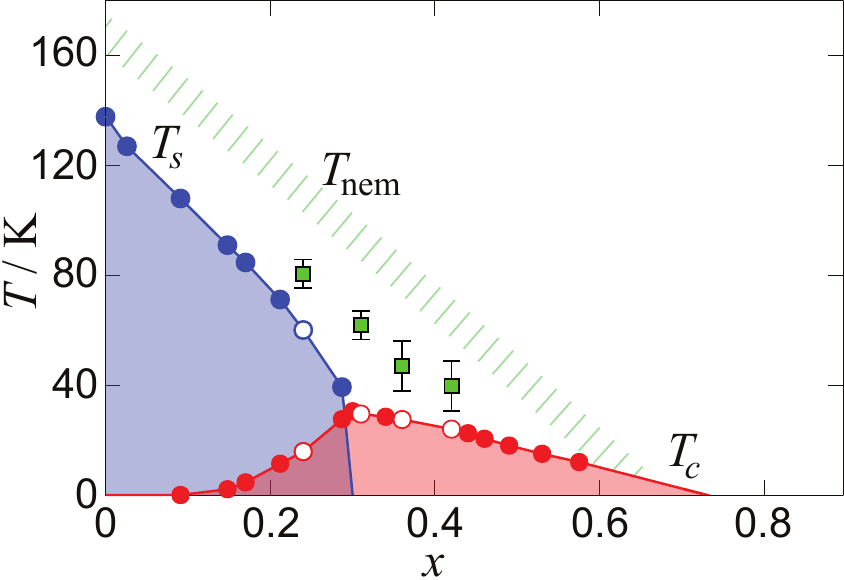}
\caption{Temperature-doping phase diagram indicating structural/magnetic (blue circles) and superconducting (red circles) phase boundaries from resistivity measurements. Green squares indicate the onset of broken C$_4$ symmetry as detected by birefringent photomodulated reflectance and open circles mark the samples studied via both techniques.  The hashed green line indicates the nematic transition observed in~\cite{kasahara2012electronic}. \label{fg:phasediagram}}
\end{figure}

Figure~\ref{fg:phasediagram} shows a phase diagram of the P:Ba122 system in the $T-x$ plane that is based on characterization of a large set of samples with $0<x<0.58$.  The circles indicate phase transitions as determined by transport and magnetic susceptibility measurements. The lines labeled $T_s$ and $T_c$ agree well with previously established boundaries of an orthorhombic phase with stripe-like antiferromagnetic order and superconductivity, respectively~\cite{allred2014coincident}.  The hashed line labeled $T_\text{nem}$ indicates where Kasahara \textit{et al.} reported the onset of C$_4$ breaking in magnetic susceptibility~\cite{kasahara2012electronic}.  Finally, the solid squares indicate where we observe the onset of birefringence in four samples selected from the larger group for optical measurements (the open circles indicate the temperature of strong features in the resistivity for these samples).

From the perspective of the overall phase diagram, the onsets of C$_4$ symmetry breaking observed by photomodulation are quite surprising, as they appear to define a line distinct from the tetragonal-to-orthorhombic phase transition.  As mentioned above, the same symmetry cannot be broken at two temperatures.  One way around this difficulty is to suppose that, while spontaneous C$_4$ symmetry breaking takes place only at $T_s(x)$, strain-induced birefringence can be observed at higher temperatures as consequence of the large nematic susceptibility. This is the conclusion reached from photomodulated reflectance data in Ba(Fe$_{1-x}$Co$_x$)$_2$As$_2$~\cite{mihailovic2}, where birefringence was also observed at temperatures well above $T_s$. In that work it was suggested that uniaxial strain could result from the combination of surface inhomogeneity in the form of terraces and laser heating. However, whereas the temperature dependence of birefringence in Ba(Fe$_{1-x}$Co$_x$)$_2$As$_2$ is consistent with a power-law susceptibility, the onsets that we observe in the P:Ba122 system appear to be much sharper.

The second possible resolution, that C$_4$ symmetry is in fact spontaneously broken at the onset of optical birefringence, requires a reinterpretation of the transition at $T_s(x)$.  As C$_4$ symmetry is already broken, this transition would correspond uniquely to the breaking of time-reversal symmetry.  The striking manifestations of anisotropy that appear at $T_s(x)$ would then be interpreted as the amplification of the preexisting C$_4$ breaking by coupling of the lattice to spin degrees of freedom.  Future experiments can distinguish these two scenarios by measuring the strain dependence of the optical birefringence, whereby a continuous phase transition would be revealed by a divergent susceptibility.

\begin{acknowledgments}
This work was supported by the Director, Office of Science, Office of Basic Energy Sciences, Materials Sciences and Engineering Division, of the U.S. Department of Energy under Contract No. DE-AC02-05CH11231. Synthesis of $BaFe_2(As_{1-x}P_x)_2$ was supported by Laboratory Directed Research and Development Program of Lawrence Berkeley National Laboratory under Contract No. DE-AC02-05CH11231.
\end{acknowledgments}

\bibliography{biblio}

\end{document}